# BMW-ROOM
# An Object-Oriented Method for ASCET

**Max Fuchs and Dieter Nazareth**
BMW AG, 80788 Munich
Maximilian.Fuchs,Dieter.Nazareth@bmw.de

**Dirk Daniel and Bernhard Rumpe**
Department of Computer Science
Munich University of Technology
Dirk_Daniel@mckinsey.com, rumpe@forsoft.de

## ABSTRACT

This paper presents an object-oriented method customized for a tool-assisted development of car software components.

Tough market conditions motivate smart software development. ASCET SD is a tool to generate target code from graphic specifications, avoiding costly programming in C. But ASCET lacks guidelines on what to do, how to do it, in what order, like a fully equipped kitchen without a cooking book. Plans to employ the tool for BMW vehicle software sparked off demand for an adequate, object-oriented real-time methodology.

We show how to scan the methodology market in order to adopt an already existing method for this purpose. The result of the adaptation of a chosen method to ASCET SD is a pragmatic version of ROOM, which we call BROOM. We present a modeling guidebook that includes process recommendations not only for the automotive sector, but for real-time software development in general.

The method suggests to produce early prototypes that are validated and refined to completion. BROOM offers phase-independent, harmonic guidelines. Product requirements, in form of scenarios, are transformed through several activities into operational models. BROOM takes advantage of ASCET's rich experimentation- and code generation features. These allow to validate emerging models on button press. The factual development of a simplified heating/cooling system at BMW serves as a running example throughout the paper.

## INTRODUCTION

The objective of this paper is to develop an object-oriented method for the CASE-tool ASCET SD [4] which is based on the ROOM-Method [8]. German car-manufacturers hold a high market share and enjoy reputation all over the world. However, fierce competition forces top players like BMW to shake off low cost manufacturers by creating innovative value to the customer. A case in point is electronic control of elementary but complex processes, such as motor control, brake control, acceleration control, and more. To keep vehicle margins on an attractive level, but to avoid imitation, control devices are built of standard hardware components with individual software. Thus emphasis lies on the development of reliable real-time software. Computer aided software engineering (CASE-) tools allow to develop software efficiently. The focus here is on a new CASE-tool called ASCET SD [4], which generates controller code from graphically specified components. Yet ASCET SD lacks suggestions on how to develop software. So there is a need to enrich ASCET SD with a methodology that offers guidelines on what to do, how to do it, in what order, and so on. As the tool itself fosters an object-oriented (OO-) approach, so should the methodology.

The remaining part of the introduction declares the notions software engineering and object-orientation. Part two shows the development of a suitable method for ASCET SD. The core of this paper is part three, where we present our method with regards to an exemplary heating/cooling system. The last part is dedicated to concluding comments.

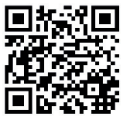

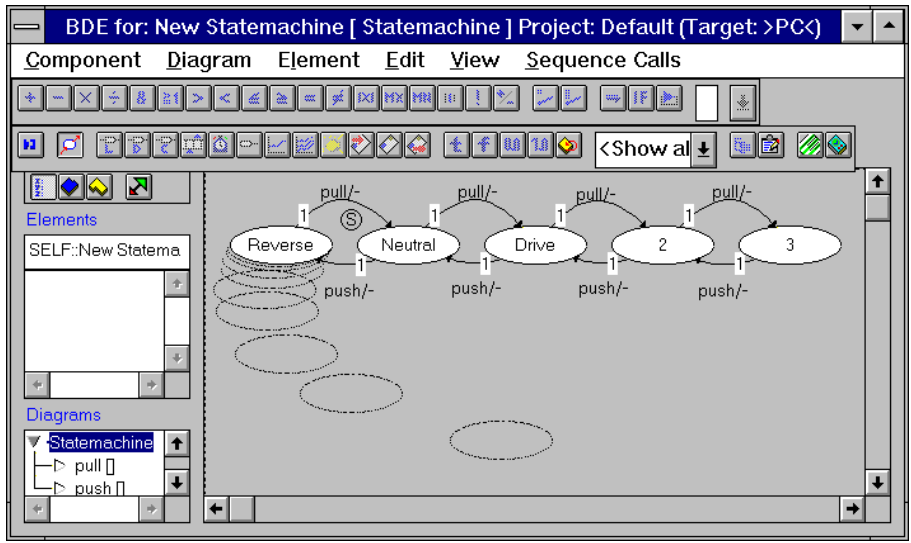

Figure 1: ASCET gear stick example

SOFTWARE ENGINEERING.

To develop software is mentally challenging due to its high complexity and immaterial existence. Problems emerge of human shortcomings, e.g. insufficient precision or the rare ability to penetrate complex facts. The scientific discussion to overcome these problems is called software engineering. The traditional software engineering process model suggests to organize software development by successive phases like a waterfall [7]: In the first phase, the software requirements are analyzed. Then the fundamental structure is designed by decomposition into easy to model components. These components are implemented and tested individually before being integrated and tested as a whole. The last phase is dedicated to the maintenance of the installed software.

It is an accepted fact that this model has disadvantages, above all a missing feedback to prior phases. Especially errors in analysis and design can only be corrected at high cost and threaten the whole project if discovered not before test. In the course of time, improved or new models have emerged, among them prototyping [2] and the spiral model [1]. Both foster an incremental approach. This means creating a meager but executable prototype that reveals deficiencies at an early stage. This information is used to construct an improved prototype, which can be verified again, and so on. This evolutionary prototyping cycle is repeated until the software is finished or a shortage in time or money allows to deliver at least a version with limited capabilities. These considerations form a foundation for the new method for ASCET.

OBJECT-ORIENTATION.

New tools and process models require new software development methods. The idea of object-orientation (OO) has become popular lately. Object-orientation models a system analogous to nature as objects that interact and thus trigger certain patterns of behavior. Objects incorporate inheritable attributes (data) and methods (functions or procedures). Each object is uniquely named by its identifier and objects can be dynamically created as instances of their classes.

Persons, things of the real world, or things of imagination may be objects. Objects incorporate both structure and behavior. Structure means that objects may contain attributes that store the object's data. Often structure means also that objects contain other objects. Structure is used to express the static properties of an object. In contrast, behavior is used to express the dynamic properties of an object. Often, behavior is specified by finite state machine diagrams. These give a visual impression of the object's life-cycle. The idea of encapsulation requires behavior and structure to be melted together into an information hiding unit, which can be accessed by other objects only through an explicit interface.

The most discussed but not most important property of objects is inheritance. Inheritance in computer science works similar to inheritance in nature. In nature, children inherit structure (e.g. skeleton, muscles, …) and behavior (e.g. habits, preferences, …) from their parents. In computer science, objects inherit structure (e.g. methods, attributes, …) and behavior (e.g. a state machine diagram) from their parent objects. The inheritance mechanism allows to save money and time when developing new or evolved products.

**METHOD DEVELOPMENT**

This part firstly describes ASCET SD, which is the tool for which the method shall be developed. Furthermore, it shows how to compare, evaluate and choose a suitable method for the OO-market. Finally , it reports on the adaptation of the chosen method to ASCET SD.

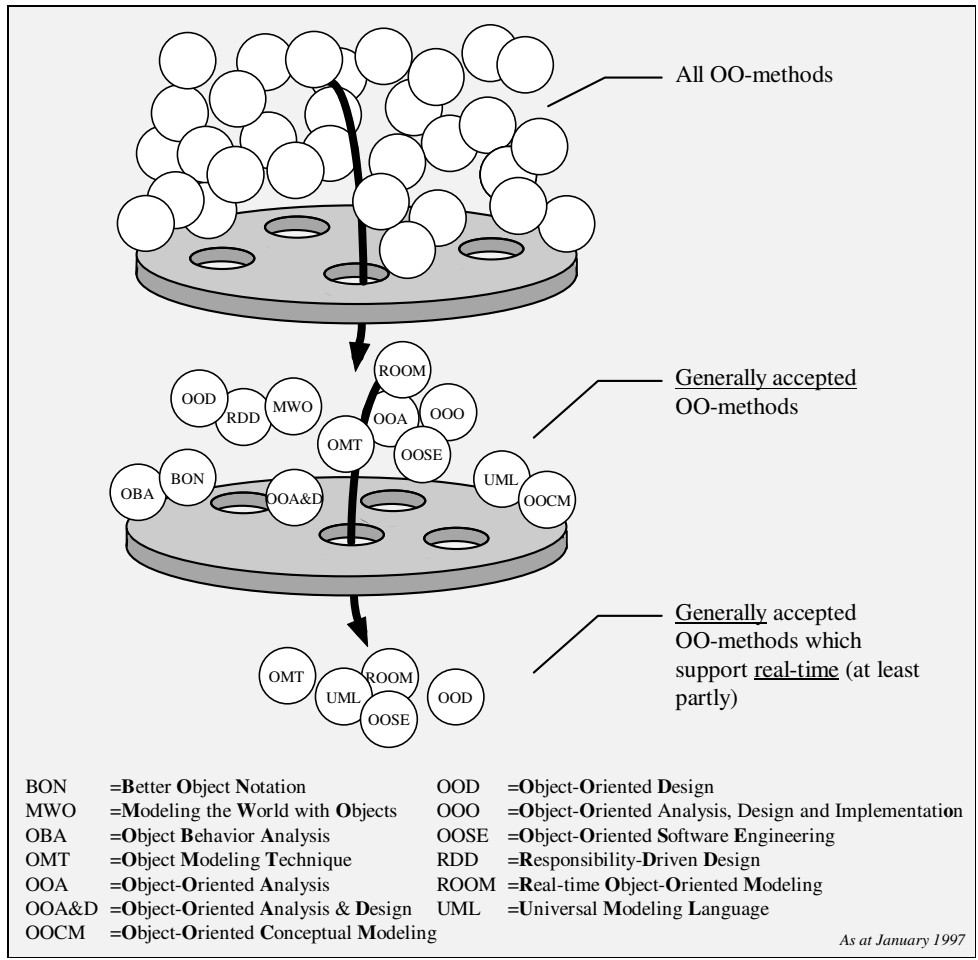

Figure 2: OO-method market analysis

WHAT IS ASCET SD?

ASCET SD (short: ASCET) is a tool intended to substitute costly C coding. The tool offers rich simulation- and prototyping facilities as well as target code generation on button press. Target code is code for the vehicle's processors (micro-controllers). Actual programming is done by drawing high level diagrams, which are transformed automatically into the desired low level code. To be specific, ASCET is a class-based (i.e. object-oriented, but without inheritance) development environment for embedded real-time software. Real-time software is software that has to respond within given time limits. In general, control systems are hybrid real-time systems. In this context, a hybrid system is a system that incorporates both continuous and discrete portions. Continuous portions are used to model physical aspects of the real world with continuous character in time (constantly busy), e.g. a temperature sensor. Discrete portions, in contrast, are used to model physical aspects of the real world with discrete character in time (sporadically busy), e.g. a vehicle's gear-stick. ASCET supports the development of hybrid real-time systems. See Figure 1 for an exemplary screen shot of ASCET with respect to the gear-stick example.

HOW TO ADOPT A METHOD

Generally speaking, it is sensible to check out other people's work before inventing something completely new; especially if these people are experts on that matter. Concerning OO-methods, much research has been done already, so there is a good chance to save a lot of money and time if a mature OO-method can be adopted to the ASCET tool.

Of all OO-methods available on market, just generally accepted methods with (at least partial) real-time support are candidates for adoption to ASCET. Generally accepted methods posses approved quality, they are publicly accessible and, last but not least, they raise the acceptance or our work. Real-time methods care for the specific needs of controller software development. See Figure 2 for the coarse selection process, which yields five possible candidates.

These candidates require detailed comparison. An OO-method consists mainly of two important aspects: Notation and process issues.

- Notation is a textual and/or graphic formalism to represent the basic, the structural, the dynamic, and the real-time properties of the model under

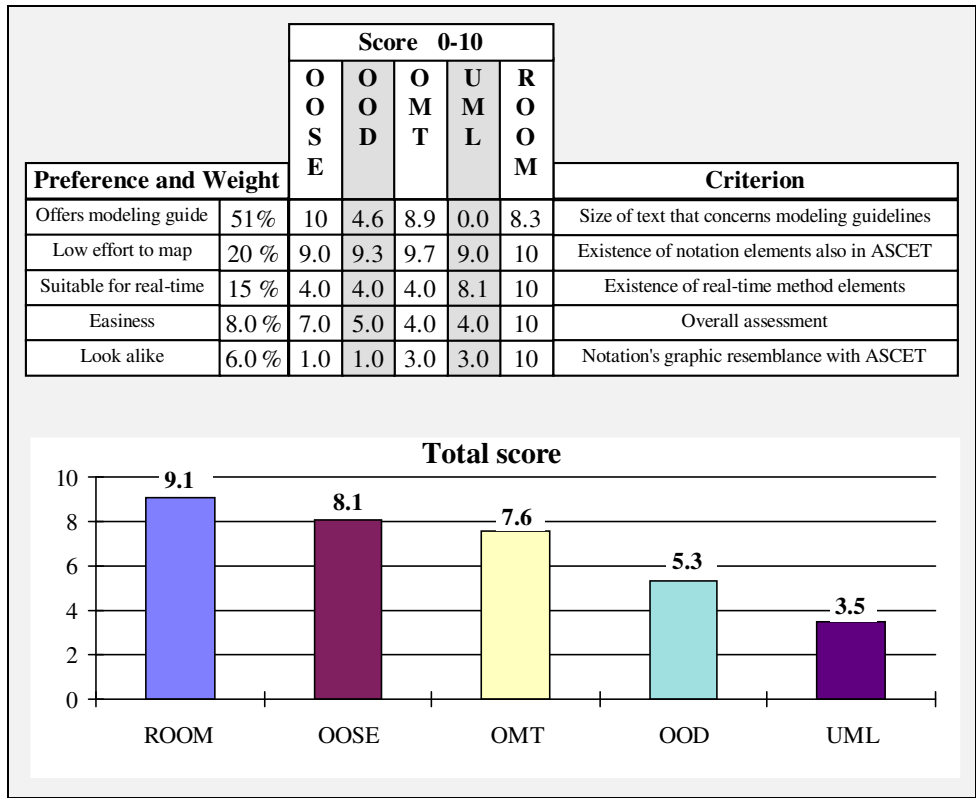

Figure 3: Method adoption scoring model

- development. Here, a model is a simplified image of reality that acts as a skeleton of the desired piece of software.
- Process issues concern people developing software; i.e. how to employ the method's notation the right way, how to develop a model and how to manage the software development organization.

The point of interest is: To which degree do the creators of the methods support our objective? In order to find a suitable method, it is necessary to agree on preferences. The most important preference in this case is that the adopted method offers modeling guidelines, because this is exactly the missing point in ASCET; this preference is more important than the sum of all other preferences. Furthermore, it must be possible to map the results of the method to the ASCET tool and to treat real-time problems. Easiness and graphic resemblance are of lower importance. With the help of a scoring model, the final decision can be made. Scoring models have been developed to find optimal solutions for qualitative decision problems. Figure 3 shows the results including the winner of the contest: The ROOM method.

CUSTOMIZATION

The Real-time Object-Oriented Modeling methodology (ROOM method) [4] is an OO-method that is explicitly dedicated to distributed event-driven real-time applications. Its is the result of more than ten years design experience in the telecommunication industry. The ROOM development process is an incremental and iterative loop with continuous feedback. It produces early operational models that are refined to implementation. ROOM models can be compiled and executed with a commercial CASE toolset (ObjecTime) at any stage of development. From analysis through design to implementation, ROOM uses phase-independent, continuous concepts that avoid discontinuities and semantic gaps between different model representations.

Since ROOM was neither developed for the specific demands of the vehicle industry nor for realization with the ASCET tool, it is evident that the ROOM method needs adaptation. A notation mapping feasibility study shows that most ROOM concepts can be mapped to ASCET, except for inheritance, dynamic structure/substitution and advanced state machine concepts. According to the developers of ASCET, inheritance and advanced state machine concepts will be implemented in near future. The adaptation of ROOM for our purpose is what we call BMW-ROOM, or BROOM for short. To put it in a nutshell, BROOM is a simplified version of ROOM with specific customizations for small size processors. For example: ROOM uses a queued messages for communication between objects. This mechanism is replaced in BROOM to a large degree by method calls for performance reasons.

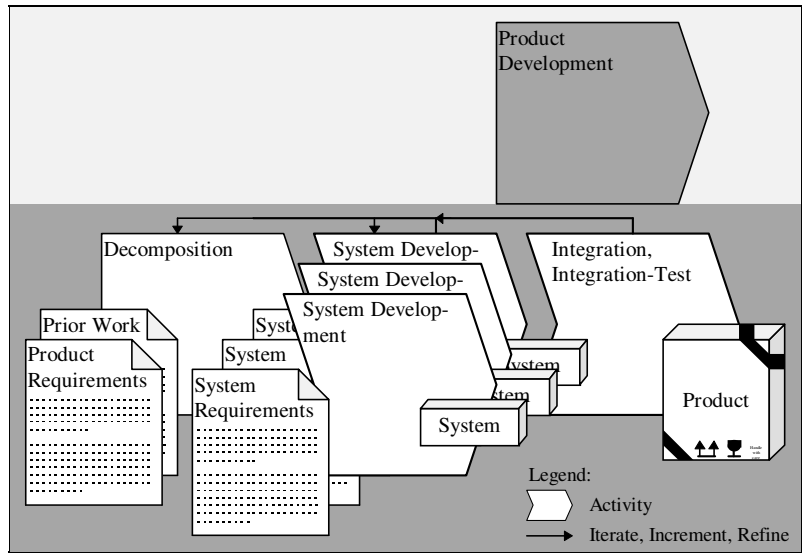

Figure 4: Product Development

Regarding notation, BROOM builds on ROOM, but includes also some language elements of ASCET, such as certain connecting lines between objects. Regarding process issues, BROOM is similar to ROOM, but includes also some of the best ideas of other OO-methodologies.

**BROOM**

This part presents our method, BROOM. To illustrate the points of interest, we provide a running example. It represents the factual development of a simplified heating/cooling system performed by BMW.

BROOM suggests to do much of the development in mind and on paper first, before it comes to programming. This is because we believe that high-level concepts should be developed and discussed in a team (-room), and not individually with a drawing utility in front of a computer. Why should software developed in a team? Above all due to the complexity of software, which can only be handled by the strategy of divide-and-conquer. This means for our purpose: Splitting the product recursively into systems and assigning people to each system. Since these systems must cooperate, so should their developers. For example, it is necessary that the developers of air conditions consult people busy with power supply.

The software in development is usually part of a product. For example, software to control brakes is sold to customers together with a vehicle, not alone. So software development is part of product development. Product development is a process that transforms product requirements into products. In addition, many products must be installed, maintained and more; machinery, for example. This guide does not focus on installation, maintenance and evolution. Product requirements describe informally what the product must do; not how. As for the running example, the product requirements just state that the product (i.e. the vehicle) should feature an heating/cooling system which controls inside temperature. It does not state how the system is intended to work. Often, requirements appear as scenarios. A scenario is a little history that describes, by actions or sequences of actions, how something is used or what it must do. The product requirements stem from marketing research or directly from the customer, often through interviews. They state the scope of the problem, priorities to development, related resources, assumptions, and especially in control engineering, specifications concerning performance. In general, product requirements tend to be ambiguous, inconsistent, incomplete and inadequate.

The development of a product in a piece is too complex. Instead, it is recommended to decompose the intended product into smaller chunks called systems, see Figure 4. These systems can be developed with less effort, often in parallel. For a car, it is evident that different people are busy with different systems in parallel to reduce time to market. After system development, the systems are integrated and tested as an entire product. This guide does not focus on integration and integration-test.

A system is an encapsulated unit that serves a certain purpose. The product itself can be seen as a system, too. If systems are still to complex to handle, then they are decomposed into further subsystems, until they are small enough to be developed by a single person or a team. Decomposing a system involves defining a system boundary. The system boundary is a demarcation between the system itself and the environment of the system. Each system of the product should have a clearly defined purpose, which can be described in a few words. As for the heating/cooling system, an aggregate to heat or cool is obviously within the system. The software to control the system is also within the system. In contrast, the vehicle's inner air is not within the system. It belongs to the environment because it has no purpose (in this context). If hat hierarchic structure is not apparent, it might help to paint a circle representing each system on a sheet of paper; the distance of two circles should express the relation of the corresponding systems (small distance = strong relation). Then the structure can be composed hierarchically by combining neighbored circles according to Figure 5. This technique supports the modular approach.

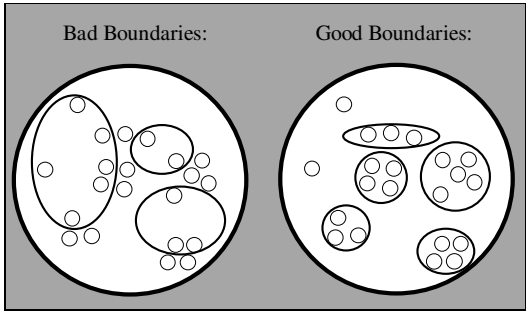

Figure 5: System boundaries

External to the system is the environment. If no idea comes to mind of what the environment of a system is, then it is useful to start thinking of an entity called 'rest of the universe'. In the course of time, more and more items of the environment may be eliminated, if they have no relation to the system. This means that the system boundary may shrink with progress in understanding of the system. The result of decomposition is a collection of system requirements; one document for each system. To generate a system requirements document for a specific system, all information of the product requirements that inspired to identify this system is gathered. This information is then combined with a system boundary proposal, which reveals what items should be in the system, and what items represent the environment. Instead of developing a system as an isolated unit, we suggest to develop a model. A model is a simplified image of reality: It includes not only the system, but also simulated environment. The document needed for model development is called the model requirements document. The model requirements incorporate the system requirements, enriched with additional requirements for simulating the environment. As for the heating/cooling system, five scenarios and the system boundary form the model requirements. See Figure 6. (Previous product- and system requirements are omitted here to avoid repetition.)

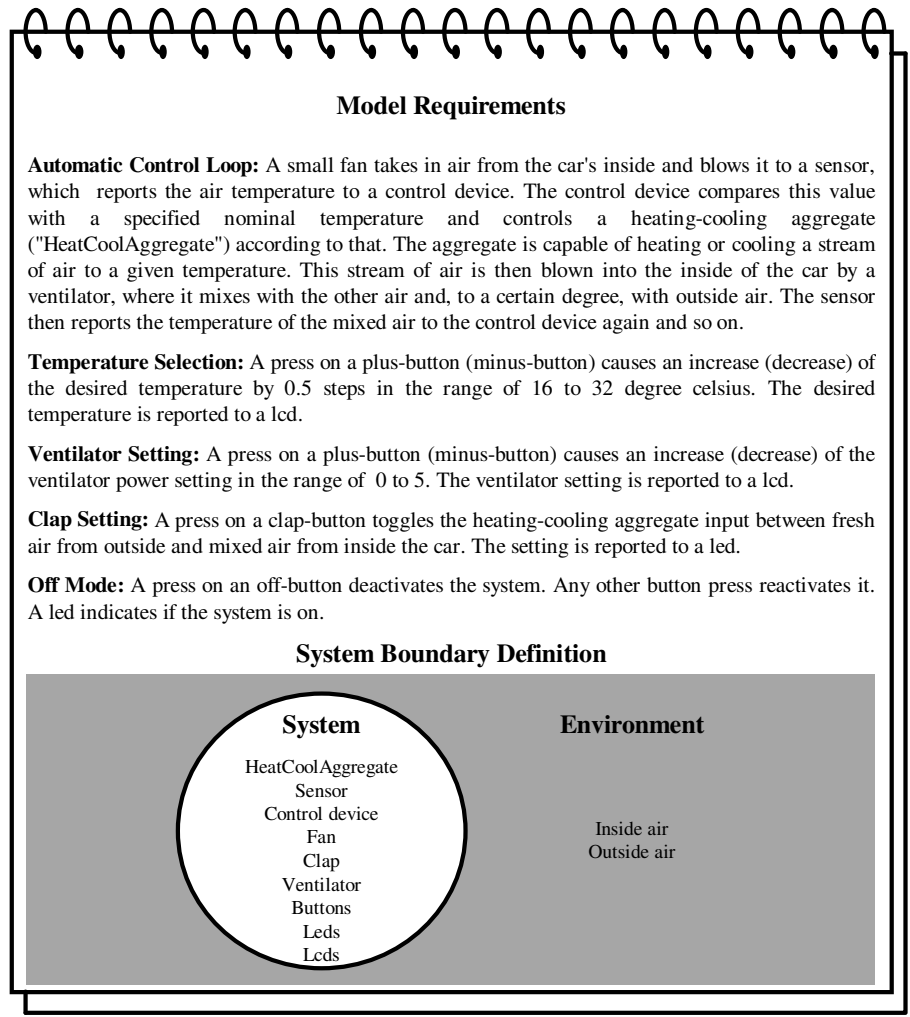

Figure 6: Model requirements

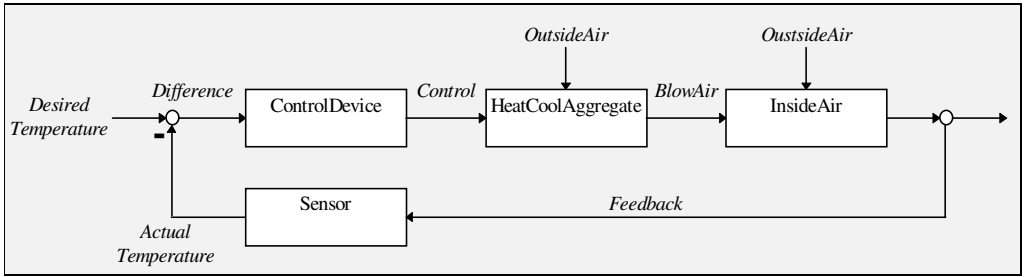

Figure 7: Control loop

The first scenario reflects a typical closed control loop, to which the theory of automatic control systems can been applied, see Figure 7.

Temperature control is a continuous process, which compares the desired temperature with the actual temperature. The actual temperature is measured by a sensor. The control device reacts and controls the effector of the model, the HeatCoolAggregate. The HeatCoolAggregate is capable of influencing the InsideAir. The trouble making point is the OutsideAir, which disturbs the heating/cooling-mechanism of the aggregate and the temperature inside the vehicle (for example through an open window).

This is a standard control problem to which standard solutions exist. According to automatic control systems theory, the HeatCoolAggregate and InsideAir can both be modeled by a so called P-T1 element. The chain of both elements shows P-T2 behavior. To control a P-T2 element, PI-controllers have proven to be of value. The ASCET library provides various types of predefined components, among them P-T1 components and PI components. The model is developed on paper first to express high-level issues and implemented later. See Figure 8.

Generally speaking, the definition of the model requirements contains modeling objectives, scenarios and message sequence charts. While modeling the modeling objectives express different aspects of software development. During project begin, the focus is set on the discovery of the basic behavior of the system and on interaction with environment. The motivation is to understand the system ('analysis'). Design alternatives are not topical at this point. Later, it is suggested to concentrate more on invention of system components ('design'). But it is important to notice at this point that discovery and invention are not subsequent phases, but interchanging activities during the entire modeling process.

A scenario relates, by an action or a sequence of actions, how something is used or what it must do. The definition of scenarios involves gathering scenarios from the system requirements and composing additional scenarios. The additional scenarios care about environment simulation. Sometimes scenarios can get very complex. As a help, complex scenarios are translated from the textual representation into a message sequence chart. A message sequence chart (MSC) is a graphic representation of a scenario; cf. [5] p. 142ff or [6] p. 86ff. It specifies the temporal order of message exchange between entities, objects, persons, and so on. The first scenario of the heating/cooling model seems to be complex, so it is transformed into a MSC. The entities are represented by vertical bars, proposed method calls are represented by arrows which point to the 'owner' of a method. See Figure 9.

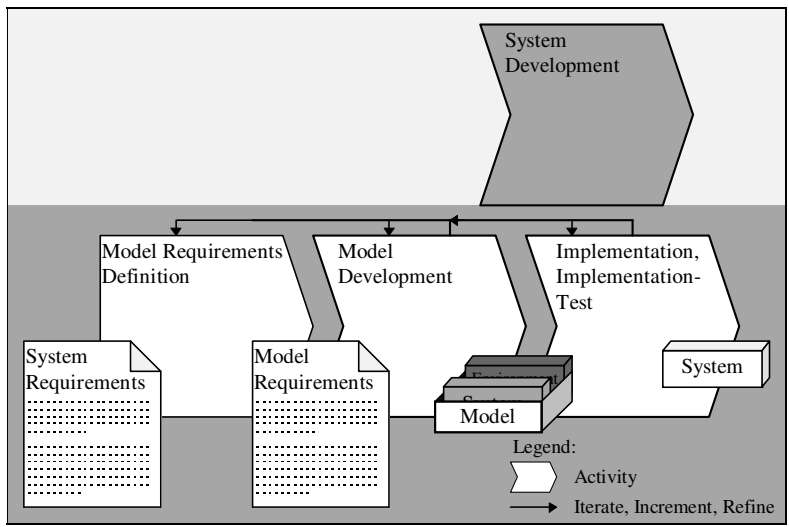

Figure 8: System development

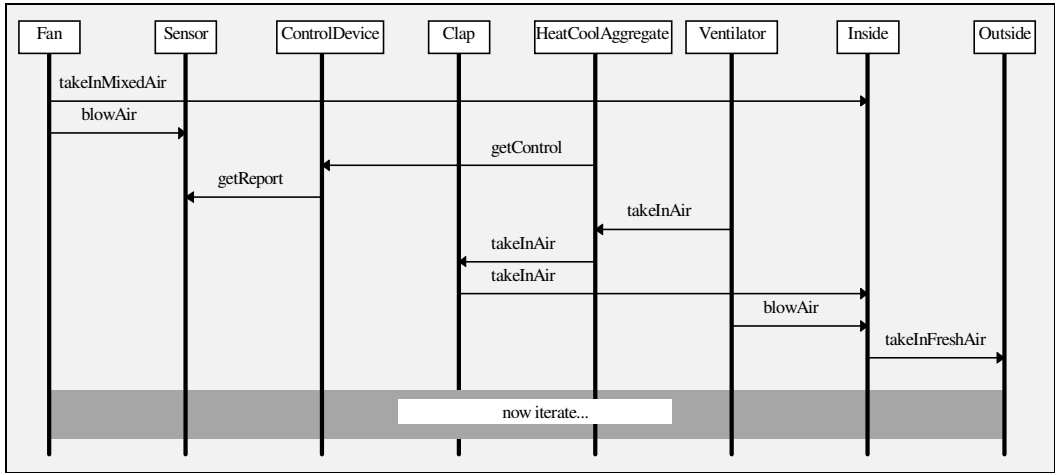

Figure 9: MSC for the control loop

Model development starts with scenario selection. A model is developed to satisfy all chosen scenarios as well as the scenarios that emerge during development and validation. However, to develop a model in one go is not suggested. Instead, one should construct prototypes that fulfill only a portion of all scenarios ('scenario package'). Prototypes are validated at an early stage and refined incrementally to completion, see Figure 10. This way, errors in scenarios or models can be detected before running into a vicious circle of redesign, time pressure, more errors, more redesign, more time pressure, ...

At first, those scenarios are chosen that do not concern the system internally, but which concern the interaction between system and environment. These scenarios are used to create a new scenario. This scenario will be called the top-level scenario. It involves few entities, often just the system and the environment. This scenario should express a typical sequence of actions that crosses the system boundary and which seems to be important. Then more scenarios are selected to obtain a new package of scenarios. The package should be small enough to achieve the goal of early validation. At the beginning of a project, it is recommended to select particularly those scenarios that decide if the project is feasible or not. If a project must be aborted, then it should be aborted early. Then scenarios are selected that resemble work of prior projects. Maybe it is possible to find reusable components that already do the job. Then scenarios are selected that represent key product properties and least-understood features. Extra-features should be concerned not before end. To sum it up: Hard work first. The model is evolved through construction and validation. If additional scenarios emerge or existing ones are modified, it is necessary to confirm that the scenarios do not contradict each other and that the model still satisfies all scenarios. The construction of a model is an interplay between three activities: Element identification, consolidation, and behavior capture.

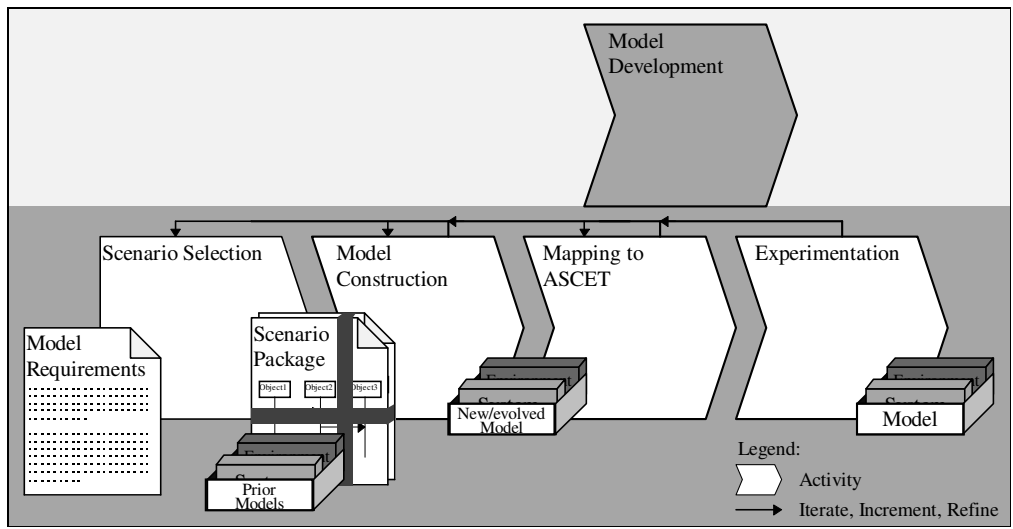

Figure 10: Model Development

Element identification

The process of element identification reveals building blocks (elements) for the model. The most important source of information is the scenario package. The output is a list of possible candidates. Elements are actor classes (including messages and methods), and data classes. An actor class is a pattern to create similar actor objects. An actor object ('actor') is an independent, concurrently active object in terms of object-orientation. In general, it will contain other actors hierarchically and it will also contain a state machine and data objects. The state machine describes the dynamic behavior of the actor. Actors are used to model components which are thought of being inherently 'busy', e.g. a controller. An data class is a pattern to create similar data object. A data object is a passive form of an actor (i.e. not busy). It is included into other actors as an encapsulated data container or utilized for information transport. In general, it contains primitive ASCET data types, other data objects, and methods to access the encapsulated data. Data objects are used to model information, such as temperature, speed, fuel consumption, and so on. Messages and methods are used for communication between actors. With each pass of the element identification phase, more and more elements are discovered. How to find elements? Element candidates are mainly entities on each side of the system boundary, described in the model requirements document, and items of the MSCs. Especially in control engineering, one will always come back to actor classes that measure some property of the environment (sensors), actor classes that compute on this information (controllers) and actor classes that influence the environment according to the corresponding results (effectors). Another technique for element identification is scanning for nouns, verbs and adjectives in the scenarios. Here are some more ideas possible candidates: Real-time purpose elements (timers, stop watches, samplers, ...), hardware elements (e.g. machines, electronic components, buildings), abstractions of physical, chemical, ... objects and - processes, connectors, coordinators, converters, negotiators, administrators, user interface objects, containers and, last but not least, persons/users.

Consolidation

The list of candidate elements usually needs to be consolidated. Consolidation involves regrouping element relations, uniting similar elements, modifying elements, and creating new elements. Three types of relations play an important role in this context: Containment, communication and inheritance. Containment is the property of objects to be part of other objects. In general, each included object is an individual, local copy. Communication allows objects to collaborate with other objects: To use each other services, to control other objects, and more. Actors can communicate with other actors by sending and receiving messages, or by calling methods or other actors and getting called. It is important to notice that actors do not address other actors directly; instead, they use special gates for this purpose, called 'ports'. This concept allows actors to be to employed in combination with various other actors, thus supporting modularity. Inheritance is a relation between actor classes and data classes, which expresses that items from the superclass are passed on to the subclass. A subclass may only have one superclass. At the moment, only structural inheritance is possible with ASCET, and only to a certain degree (no 'dynamic binding'). Besides regrouping element relations, other activities take place. For example, if element candidates are matched against each other and it is revealed that they share (nearly) identical purposes, then they are subsumed into a new element, eliminating the old.

Behavior Capture

The process of behavior capture brings the static actors to life. This is done by transforming the scenarios into state machine diagrams. A state machine diagram is a visual specification of a state machine. A state machine consists of several states and transitions. Not every actor possesses individual behavior (e.g. containers). Care has to be taken that the model meets given real-time requirements. Real-time systems show the properties of timeliness, reactiveness, concurrency, and distribution. The property of timeliness is the most important aspect for real-time models. It is necessary to validate that each actor performs its function on time. The property of reactiveness addresses the continuous interaction of actors with the environment ('event-driven system'). The environment initiates events, to which the system must react by generating output. In general, these events are of unpredictable nature. The property of concurrency concerns the intrinsic parallelism among actors. If some actor appears in several scenarios, then this hints to a scenario conflict. Often it is necessary to look at many scenarios in parallel while modeling. The property of distribution addresses the fact that some components of the systems may not be located on the same unit of computation, but may be located on other units.

As for the heating/cooling model, the first scenario is chosen to construct an early prototype with limited functionality (top-level model). To keep the model simple at first, the three activities element identification, consolidation and behavior capture are performed only in brief. The top-level model only features three actor classes: The Heat/CoolModel, Heat/CoolSystem, InsideAir, and OutsideAir. Figure 11 shows the definition of the actor class Heat/CoolModel (thick box = definition). The class contains three actor objects: Heat/CoolSystem, InsideAir and OutsideAir (thin box = reference). The stubs at the border represent communication ports, the textual descriptions next to them represent methods contained in an actor. The thin lines connect stubs of different actors, thus enabling communication through a static channel.

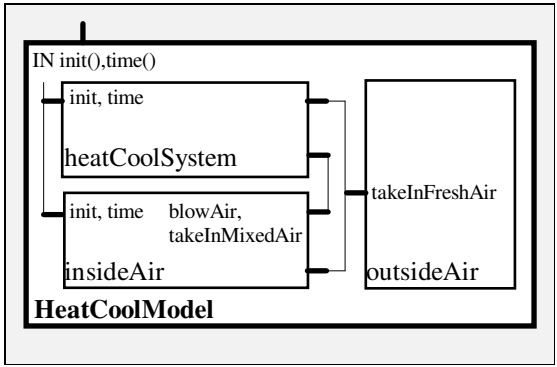

Figure 11: Top-level model

Early validation of operational prototypes is the key strategy to prevent running into dead end roads. Model validation involves mapping to ASCET and makes use of the experimentation capabilities of ASCET. After the mapping process, code is created for a personal computer (off-line experimentation) or for a real-time environment, e.g. a transputer (on-line experimentation). Of course, experimentation can never prove the complete absence of defects, only their presence. Experimentation is started with data classes, since actors rely on data classes. Then actors are tested. The primary technique to verify that actors behave according to the scenarios is to rehearse these scenarios. At first, only a single scenario from the scenario package is selected. It is necessary to confirm all actors compute correct results and perform on time. To receive realistic results concerning time, on-line experimentation is used. Later, actors are exposed to several related scenarios and it is validated that actors behave correct with regard to scenario conflicts.

If it is obvious that the model is correct, which is seldom the case in the beginning, then it is useful to freeze the current knowledge of the system using comments and reports. If the model is incorrect, then it might be sensible to confirm that no mistakes occurred during validation itself during mapping to ASCET. To find modeling errors it is recommended to apply the strategy of bisection: That means not to debug a large portion of computation in a piece, but to find a point 'in the middle', identify the area of incorrect computation, then iterate with another bisection, until the trouble spot is identified. When debugging behavior, ASCET's state machine animation is very useful. If a model satisfies the scenarios but behaves strange nevertheless, then this might hint to deficiencies in the model requirements. The development of a model is an interplay of discovery and invention. Often, the invention (design) of solutions for a model is interrupted by the discovery of the need to change or add requirements (analysis). This effect is intended and will enlarge the understanding of the system. This is the point when to go back to the requirements definition phase.

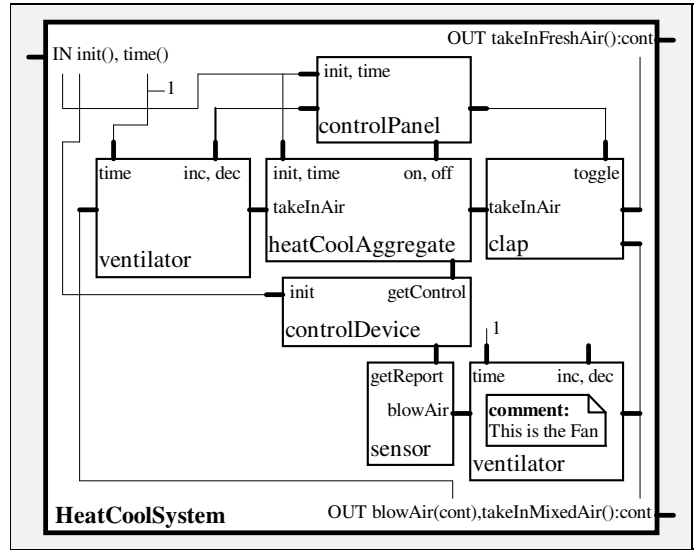

Figure 12: HealCoolSystem Actor

After validation of the top-level version of the Heat/CoolModel, more and more scenarios are taken into account in order to evolve the desired piece of software. This means getting more into the details of the actor class HeatCoolSystem. During element identification, most of the element candidates found are typical hardware abstractions. During consolidation, it is discovered that the actor class Fan is just a Ventilator class without speed control. Behavior capture is omitted here. See Figure 12 for the refined actor Heat/CoolSystem. The definition of this actor represents the closed control loop. The evolved model is again mapped to ASCET and validated. In this pass, all major properties of the desired system can be verified with the exception of the user interface features. In a last iteration of the modeling process, the model is refined to completion by filling in the remaining button, leds, displays, ... into the actor ControlPanel. The definition of these and other classes, among them classes with state machines, is omitted here. A last validation confirms that the Heat/CoolModel behaves correctly.

**CONCLUSION**

We presented a real-time software development method, customized for a CASE-tool. The method actually was intended to improve work at BMW [3]. The resulting code has proven to be efficient. Nevertheless it is still a challenge to convince software developers of the benefits of prototyping approach. It is especially hard to motivate people to develop high-level concepts in a team on whiteboard or on paper before rushing to code. Tool support for these activities is desired in order to avoid the process of mapping models manually to ASCET. Another challenge is to convince people not only of BROOM, but also of the object-oriented approach itself. Many developers got used to thinking in data flows and functions and find it difficult to shift paradigms towards object-orientation.

We do want to draw attention to the fact that BROOM is not fully developed so far. To be specific, it shows only a small potion of all development activities. On the one hand, it does not comment on hardware connect, implementation, installation, and so on. On the other hand, it lacks theoretic foundation as well as economic evaluation of benefits and costs.

But we are eager to take advantage of new ideas. We think that advanced tools and methods are essential if we want to deliver products that fascinate our customers (and yes, we want that).